\documentclass[twocolumn]{aastex63}

\usepackage{hyperref}
\usepackage{multirow}

\received{April 1, 2021}

\shorttitle{Taylor Swift}
\shortauthors{Mansfield \& Seligman}

\begin{document}

\title{I Knew You Were Trouble: Emotional Trends in the Repertoire  of Taylor Swift}

\correspondingauthor{Megan Mansfield}
\email{meganmansfield@uchicago.edu}

\author{Megan Mansfield}
\affiliation{Department of Geophysical Sciences, University of Chicago, 5734 S. Ellis Avenue, Chicago, IL 60637, USA}

\author{Darryl Seligman}
\affiliation{Department of Geophysical Sciences, University of Chicago, 5734 S. Ellis Avenue, Chicago, IL 60637, USA}

\begin{abstract}

As a modern musician and cultural icon, Taylor Swift has earned  worldwide acclaim via pieces which predominantly draw upon the complex dynamics of personal and interpersonal experiences. Here we show, for the first time, how  Swift's lyrical and melodic structure have evolved in their representation of emotions over a timescale of $\tau\sim14$~yr. Previous progress on this topic has been challenging based on the sheer volume of the relevant discography, and that uniquely identifying  a  song that optimally describes a hypothetical emotional state represents a multi-dimensional and complex task. To quantify the emotional state of a song, we separate the criteria into the level of optimism ($H$) and the strength of commitment to a relationship ($R$), based on lyrics and chordal tones. We apply these criteria to a set of 149 pieces spanning almost the entire repertoire. We find an overall trend toward positive emotions in stronger relationships, with a best-fit linear relationship of $R=0.642^{+0.086}_{-0.053}H-1.74^{+0.39}_{-0.29}$. We find no significant trends in mean happiness ($H$) within individual albums over time. The mean relationship score ($R$) shows trends which we  speculate may be due to age and the global pandemic. We provide tentative indications that partners with blue eyes and/or bad reputations may lead to overall less positive emotions, while those with green or indigo-colored eyes may produce more positive emotions and stronger relationships. However, we stress that these trends are based on small sample sizes, and more data are necessary to validate them. Finally, we present the \texttt{taylorswift} python package\footnote{\href{https://github.com/meganmansfield/taylorswift}{taylorswift} code is available on GitHub} which can be used to optimize song selection according to a specific mood.

\end{abstract}

\keywords{music - pop; music - country; artists, individual - Taylor Swift}

\section{Introduction}
\label{sec:intro}

Over her tenure as a well-known pop and country singer, Taylor Swift has produced an abundance of highly acclaimed music. She has received a total of 11 Grammy awards and is one of only two female solo artists to receive Album of the Year twice \citep{Grammy}. Her music has attained such popularity through its treatment of a wide variety of interpersonal relationships. However, as yet there have been no studies to look for overarching trends across her entire repertoire.

\citet{Seidel2016} compared Twitter interactions between researchers and celebrities, specifically focusing on how the communicative features of Twitter were used to produce status in the academic profession and popular music profession. Of particular interest, \citet{dubrofsky} performed a detailed analysis of the music video for \textit{Shake It Off} and concluded that her performance and behavior was marked by self-reflexivity, and was not representative of the authentic self. In this work, we attempt to constrain by a complementary approach, by analyzing the text itself of the entire repertoire, which provides an independent constraint on the authentic self. 

\citet{Brown_2012} provided a complementary analysis to what we present in this paper, where they examined the content of two websites devoted to Taylor Swift fandom as opposed to the lyrics in the discography itself. Moreover, they point out that by the autobiographical nature of Swift's music, Swift poses herself as an ordinary girl that many fans can relate to. 

This work represents a novel analysis to the previous attempts to disentangle the persona perpetuated in the media and the underlying emotional state. Taking the hypothesis that the lyrical structure and overall chordal tone and structure are representative of the emotional state of the artist during the time of composition, we infer Swift's overall emotional state and strength of interpersonal relationships, using the lyrics as a proxy.  We draw from the compositions in the albums \textit{Taylor Swift} (2006), \textit{Fearless} (2008), \textit{Speak Now} (2010), \textit{Red} (2012), \textit{1989} (2014), \textit{Reputation} (2017), \textit{Lover} (2019), \textit{folklore} (2020), and \textit{evermore} (2020). These albums collectively span 14 years of composition. We describe our methods of analysis in Section~\ref{sec:methods} and present our results in Section~\ref{sec:results}. Finally, in Section~\ref{sec:conclude}, we briefly describe the song-and-mood-matching python package \texttt{taylorswift} and propose directions for future research.  

\section{Methods}
\label{sec:methods}
We began by compiling a list of all Taylor Swift songs that have been written and released during her career. Before beginning analysis, three sets of songs were excluded from the sample. First, covers of songs by other artists, including Christmas music, were excluded because they convey the feelings of the original author, and therefore do not provide a valid representation of Swift's own feelings. Second, we did not analyze any songs that were written for commission, including songs written for specific events or fundraisers (i.e. \textit{Ronan}, written for a cancer charity event) and songs written for film soundtracks (i.e. \textit{Eyes Open}, written for the Hunger Games movie soundtrack). Third, songs that were collaborations with other artists were excluded when the other artist, and not Taylor Swift herself, was the primary songwriter. We assumed that the artist listed first in the song credits was the primary songwriter, while the other person primarily provided backup vocals. Therefore, all songs listed as ``Taylor Swift feat. other artist'' were included in our analysis, while songs listed as ``other artist feat. Taylor Swift'' were excluded. This division was validated by the fact that, for the large majority of dual-artist songs, the artist listed first sings most of the verses and the primary melody in the chorus, while the artist listed second sings for a smaller percentage of the verses and sings the harmony in the chorus. For example, \textit{Everything Has Changed} was written by Taylor Swift feat. Ed Sheeran and features Taylor Swift singing the melody while Ed Sheeran sings the harmony, so it was analyzed. However, in \textit{Highway Don't Care} by Tim McGraw feat. Taylor Swift, Tim McGraw sings the primary melody and most of the verses while Taylor Swift sings the harmony; this song was not analyzed.

While analyzing the remaining 154 songs Taylor Swift has written over the course of her career, we found one additional set of five songs which do not seem to reflect feelings about a relationship. Instead, these songs seemed to reflect what we refer to as ``nostalgia for childhood'' - they express a longing for a simpler time in Swift's youth. One example of this category is \textit{Never Grow Up}, in which Swift sings ``Oh, I don't wanna grow up / Wish I'd never grown up / It could still be simple.'' As the five songs in this category do not fit the overall trend of our data, we remove them as outliers. We note that these five songs make up only 3\% of the music written by Swift, so our analysis of the remaining songs is still statistically significant.

After compiling the complete list of songs to be analyzed, we listened to each song on \href{https://www.spotify.com/us/home/}{\texttt{Spotify}} (a total of 9 hours, 43 minutes, and 6 seconds of music, for those wondering what to do with free time during a pandemic) to assess the level of happiness and strength of the relationship being discussed. The website \href{https://www.azlyrics.com/}{\texttt{AZ Lyrics}} was used to obtain lyrics for each song so we could ensure no words were misunderstood. Additionally, for one of the grading criteria we will describe below, we used the website \href{https://tunebat.com/}{\texttt{Tunebat}} to obtain the key and beats per minute (BPM) of each song. If an official music video (a music video produced and published by \href{https://www.youtube.com/user/taylorswift}{\texttt{Taylor Swift's YouTube account}}) was available, we watched that video and included its visuals in our analysis.

\begin{figure*}
    \centering
    \includegraphics[width=\linewidth]{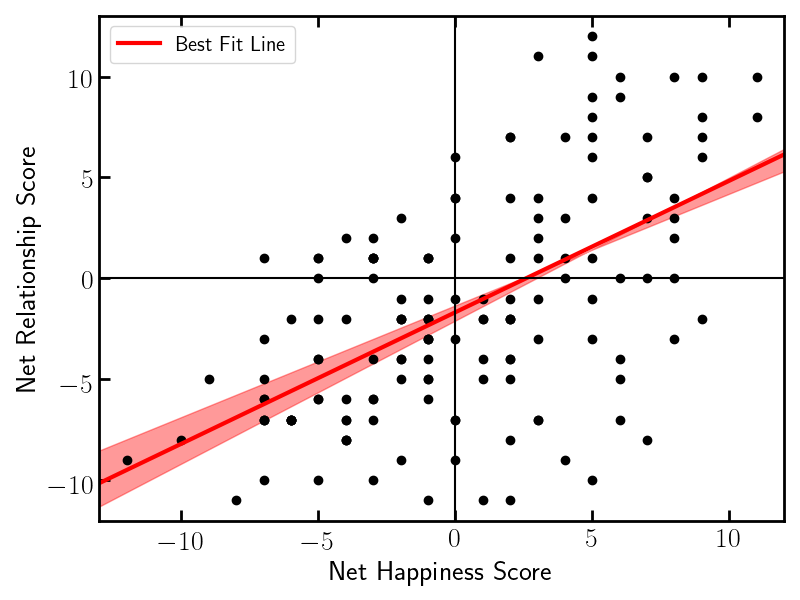}
    \caption{Scatter plot showing relationship and happiness scores for each song analyzed (black points). The red line and shaded region indicate the best linear fit to the data and $1\sigma$ uncertainties. We find a significant positive correlation between increased happiness and increased strength of a relationship.}
    \label{fig:maintrend}
\end{figure*}
Below, we describe in detail how we determined the primary object of affection (or lack thereof) in each song (Section~\ref{sec:miq}), and we lay out the point scales we used to grade each song (Sections~\ref{sec:happy} and \ref{sec:relationship}).

\subsection{Defining the ``Male in Quesiton'' (MIQ)}
\label{sec:miq}

The ``Male in Question'', or MIQ was defined to be the person in the song towards whom Taylor Swift speaks directly, or, in the absence of a second person pronoun, directs the preponderance of her feelings. Taylor Swift's feelings toward the MIQ are the topic of our full analysis here, regardless of any mention of other characters in her songs. Note that the MIQ does not necessarily have to be male: in fact, Swift breaks this trend in several of her songs. For example, in the \textit{Bad Blood} music video, Selena Gomez plays the role of the MIQ, and in the song \textit{Welcome to New York} the object of affection appears to be New York City itself and not a human at all. However, as Swift tends to write her songs from the perspective of a woman in a heterosexual relationship, we generalize and refer to the primary object of emotion in all songs as male.

In most cases, identification of the MIQ is straightforward, as only one object of affection is described in the song. However, in some instances, multiple boys make an appearance in the lyrics. In these cases, we considered the MIQ to be the one that is the primary subject of the song. The primary subject can generally be identified as the person to whom the chorus is addressed, while additional characters discussed in the verses are often secondary. For example, in the verses of \textit{Begin Again}, Taylor Swift bemoans the time spent on a previous bad relationship (``He never liked it when I wore high heels''). However, in the choruses, she focuses her attention to the positive traits of her new man (``You throw your head back laughing like a little kid''), so in this case her current beau is the MIQ. The song \textit{The Way I Loved You} flips this scenario on its head. Swift spends the verses talking about how acceptable her current beau is (``He's charming and endearing and I'm comfortable''), but the focus of the choruses is her fraught but exciting relationship with a previous man (``But I miss screaming and fighting and kissing in the rain''). Therefore, in this case we treated her previous beau as the MIQ.

\subsection{Definitions of Happiness Criteria}
\label{sec:happy}

The level of happiness or sadness Swift feels in each song was graded according to four criteria, each of which can be given a score between -3 and +3. This grading system therefore placed a quantitative number between -12 and +12 on the level of happiness in each song.

The first of these four criteria, which is described in Table~\ref{tab:selffeel}, measures how Swift feels about herself in the song. This criterion is fairly straightforward, and the examples in Table~\ref{tab:selffeel} show how it is applied to songs in practice.

The second of the four happiness criteria, which we refer to as the ``Glass Half Full'' criterion, measures Swift's outlook on life. This criterion, which is described in detail in Table~\ref{tab:glassfull} is effectively just a balance of the imagery in the song: is it primarily positive, or negative?

The third of the happiness criteria is split into two halves, as shown in Table~\ref{tab:stagegrief}. If the song mostly embodies negative emotions, the negative half of the scale determines which stage of grief best fits the song: depression, anger, bargaining, denial, or acceptance. If the song mostly embodies positive emotions, the positive half of the scale asks whether Swift is primarily focused on achieving happiness for herself only, or is working to make others happy as well.

Finally, the fourth happiness criterion measures the tempo and musical feel of the song. This criterion is not described in a table, as it is structured slightly differently than the others. As mentioned in Section~\ref{sec:methods}, we used the website \href{https://tunebat.com/}{\texttt{Tunebat}} to obtain the key and BPM of each song. For the Tempo/Musical Feel criterion, a song received +1 point for a major key and -1 point for a minor key, as in Western music major and minor keys are generally associated with ``happier'' and ``sadder'' songs, respectively. Songs also received +1 point for a fast tempo and -1 point for a slow tempo, as faster songs also generally feel happier. We generally used 100 BPM as a dividing line between fast and slow songs, although some songs near this boundary were shifted based on the feel of the song, as a song with many subdivisions at 95 BPM can often feel faster than a song without as many subdivisions at 105 BPM.

These two measurements of the song's tempo and musical feel can only produce scores between +2 and -2, however. We chose to reserve the scores of +3 and -3 for songs which had what we refer to as a ``hyperbolic line''. Hyperbolic lines are cases in which Swift suddenly increases the emotional intensity of a song incredibly rapidly. A score of +3 was given to songs where there is a sudden increase of positive emotions, such as a marriage proposal in the middle of a first date. Similarly, a score of -3 was given to songs where there is a sudden increase of negative emotions, such as a threat of violence or murder. Table~\ref{tab:hyperbole} provides a full list of the hyperbolic lines in Taylor Swift's collected works.

\subsection{Definitions of Relationship Criteria}
\label{sec:relationship}

In a similar manner as the level of happiness, the strength of Taylor Swift's relationship, (or lack thereof) in a given song was graded according to four criteria with scores on a scale ranging from -3 to +3. The relationship strength metric therefore also placed a quantitative number between -12 and +12 on each song.

The first criterion, which is described in Table~\ref{tab:seriousness}, was referred to as the ``Seriousness of Topics Discussed''. In the case of a song about a past relationship, this criterion measured the amount of animosity surrounding the breakup. In a song about a current relationship, this criterion measured how committed Swift and the MIQ were to each other based on what types of topics they discussed together. For example, discussion of marriage, death, or similarly serious issues indicates a much more serious relationship than discussion of how cute someone looks from across the room.

The second criterion, which is described in Table~\ref{tab:future}, used the song lyrics to guess at the future prospects of the relationship. The third criterion, which is described in Table~\ref{tab:malefeel}, quantifies how the MIQ feels about Taylor Swift. Both of these criteria are fairly straightforward and explained in full in Tables~\ref{tab:future} and \ref{tab:malefeel}.

The final criterion for the strength of the relationship, which is shown in Table~\ref{tab:together}, measures how much time Taylor and the MIQ seem to spend together throughout the course of the song. The grades between -2 and +2 for this criterion simply look at the lyrics as a whole and weigh whether Taylor and the MIQ spend more time apart or together throughout the song. The lowest and highest scores on this scale, however, are reserved for specific scenarios which were seen in a handful of songs. The score of +3 was reserved for songs in which, in addition to spending all her time with the MIQ, Swift seems to have no identity as an individual separate from her relationship with the MIQ. The score of -3 was reserved for songs in which there were significant barriers preventing Swift and the MIQ from spending time together. These barriers generally took one of two forms: either Swift and the MIQ are physically distant from each other (i.e., on different continents), or there is another woman/man dating one of them. The one exception to these two cases is the song \textit{no body, no crime}. In this song, Swift's character kills the MIQ - we, understandably, count this as an insurmountable barrier preventing their joint actions.

\begin{figure}
    \centering
    \includegraphics[width=\linewidth]{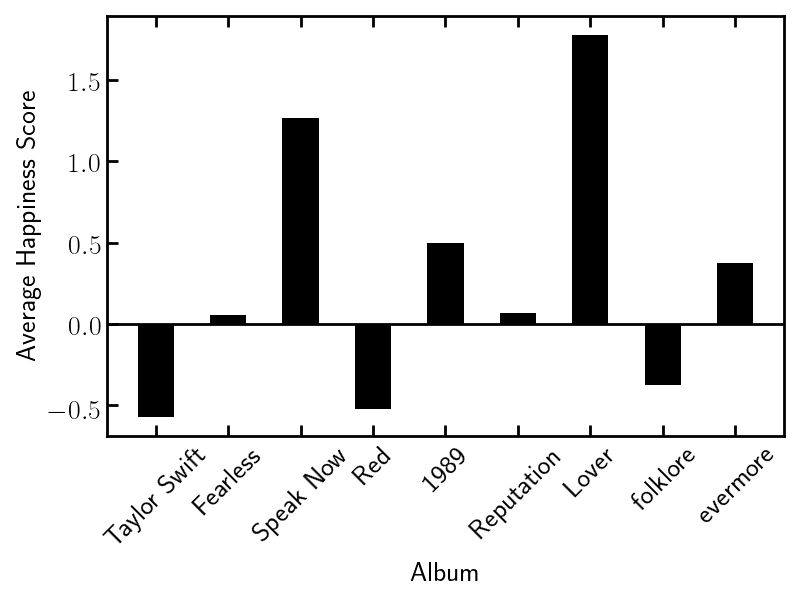}
    \caption{Mean happiness score by album, sorted by date of publication. The happiness score does not show any clear trends over time, with albums seeming to alternate between positive and negative scores.}
    \label{fig:happyalbum}
\end{figure}

\begin{figure}
    \centering
    \includegraphics[width=\linewidth]{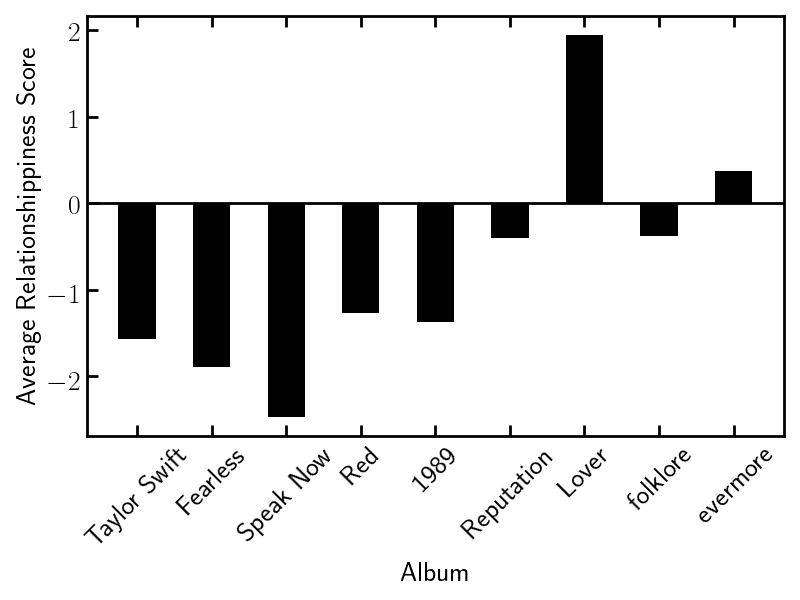}
    \caption{Mean relationship score by album, sorted by date of publication. The mean relationship score shows a trend toward lower scores through the first three albums, which then reverses and shows increasing relationship scores until a peak at the album \textit{Lover}. The two albums published in 2020 (\textit{folklore} and \textit{evermore}) show mean relationship scores stagnating near zero.}
    \label{fig:relationshipalbum}
\end{figure}

\section{Results}
\label{sec:results}

\subsection{Trends in Happiness and Relationship Scores Amongst Swift's Full Repertoire}

The primary result of our research, examining the correlation between happiness and level of devotedness to a relationship, is shown in Figure~\ref{fig:maintrend}. Note that based on the criteria defined in Section~\ref{sec:methods}, a positive score indicates net happiness or net devotion to a relationship, while a negative score indicates net sadness or net dissolution of a relationship. The data show a clear correlation, with higher happiness scores associated with higher relationship scores. In order to quantify this correlation, we performed a simple linear trend fit to the data. To determine errors on this trend, we randomly selected 70-song sub-samples of our primary data set and fit a linear trend to each sub-sample. We performed this sub-sample fitting 100 independent times. The red line and shaded area show the best-fit trend and 1$\sigma$ error bars determined by this method. We find a trend of
\begin{equation}
    R=0.642^{+0.086}_{-0.053}H-1.74^{+0.39}_{-0.29},
\end{equation}
where $R$ is the relationship score and $H$ is the happiness score.

\begin{figure}
    \centering
    \includegraphics[width=\linewidth]{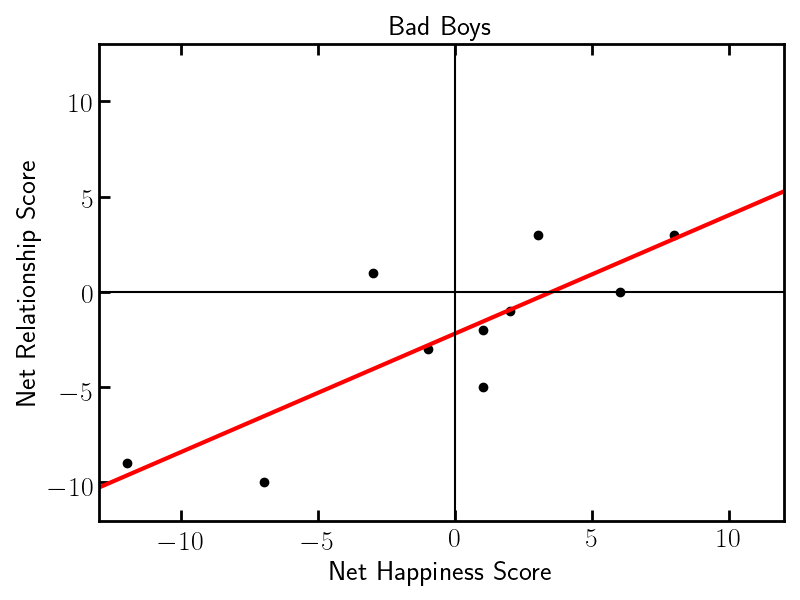}
    \caption{Relationship and happiness scores for all songs in which the MIQ is designated as a ``Bad Boy'', as described in Section~\ref{sec:subtrends}. The red line again shows a best linear fit. We find that the linear fit has a similar slope to that of the full data set, but that the intercept is $1.5\sigma$ lower, which suggests that relationships with ``Bad Boys'' are less happy overall, regardless of the strength of the relationship.}
    \label{fig:badtrend}
\end{figure}
We also examined trends in the mean happiness and relationship scores by album, which are shown in Figures~\ref{fig:happyalbum} and \ref{fig:relationshipalbum}. We find no clear trend in mean happiness score over time, with albums alternating between positive and negative mean scores seemingly randomly. However, we see more significant trends in the mean relationship score over time. The first three albums, published between 2008 and 2012, show negative mean relationship scores which consistently decrease over time. Of particular interest, the 2012 album \textit{Red} has the lowest mean score, after which point the scores consistently increase until the maximum mean relationship score is achieved in the 2019 album \textit{Lover}. The most recent albums, \textit{folklore} and \textit{evermore}, show mean relationship values near zero. We hypothesize that these trends may be due to the age of Taylor Swift at the time of writing. The albums \textit{Fearless}, \textit{Speak Now}, and \textit{Red} were written when Swift was in her late teens and early twenties, which is generally believed to represent a time  marked by short-lived and casual relationships. We hypothesize that the increase in relationship scores throughout the albums from \textit{Red} to \textit{Lover} is caused by the general increase in seriousness of relationships during mid-to-late twenties compared to teenage relationships. Finally, we hypothesize that the near-zero mean relationship scores of \textit{folklore} and \textit{evermore}, which were both published in 2020, represent the struggle we have all experienced to form new, meaningful relationships while in quarantine due to the global pandemic.

\subsection{Secondary Trends}
\label{sec:subtrends}

In addition to the primary trend shown in Figure~\ref{fig:maintrend}, we examined several subsets of the data to elucidate Swift's feelings about particular types of MIQs. First, we examine songs with MIQs who we designate as ``Bad Boys''. The ``Bad Boy'' designation is assigned to two groups of songs: first, we assign it to all songs in which Taylor Swift admits she knows the MIQ will be bad. For example, in \textit{...Ready For It?}, Swift sings ``Knew he was a killer, first time that I saw him / Wonder how many girls he had loved and left haunted''. We also assign the ``Bad Boy'' designation to songs where someone in the general community has warned Swift about the MIQ. For example, in \textit{Superman}, Swift describes her encounter with the MIQ by saying ``Something in his deep brown eyes has me sayin' / He's not all bad like his reputation''. In this case, although Taylor has no first-hand knowledge of the MIQ's behavior, she admits that she has generally heard of his bad reputation. We identified 10 songs which fit these criteria.

Figure~\ref{fig:badtrend} shows the happiness and relationship scores of all ``Bad Boy'' songs. We again fit a linear relationship to the data and found a best fit of
\begin{equation}
    R=0.622H-2.18.
\end{equation}
This fit does not show a statistically significant difference in slope when compared to our fit to the full data set, which suggests that the overall trend in increased happiness with increased strength of relationship is not affected by the MIQ's status as a ``Bad Boy''. However, we find that the intercept of this fit is $1.5\sigma$ lower than that for the full data set. While this result is not statistically significant and is based on only a small subsample of 10 songs, it suggests that in all cases Swift is less happy with a ``Bad Boy'' than with an MIQ in the general population, regardless of the strength of the relationship. If this result generalizes to more than just this specific test case, it suggests that one should always trust your own instincts and the opinions of your friends when considering a potential partner.

\begin{figure}
    \centering
    \includegraphics[width=\linewidth]{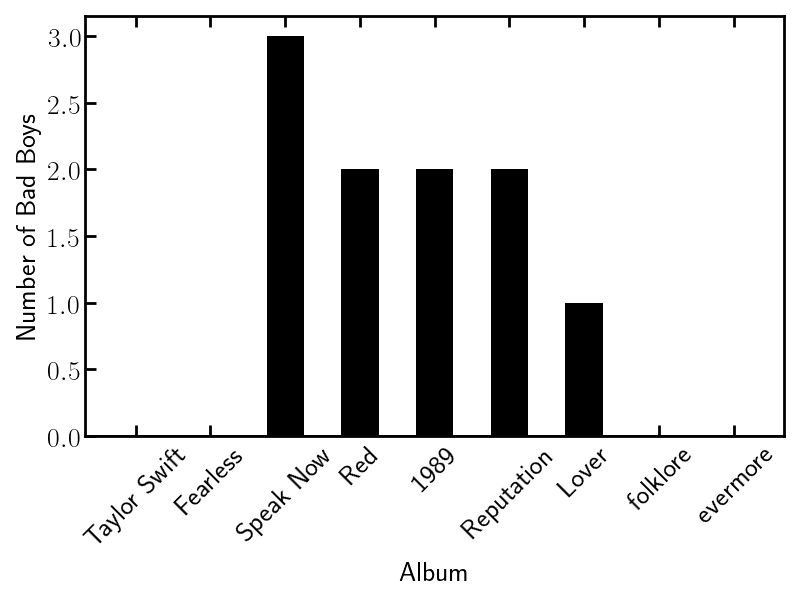}
    \caption{Number of ``Bad Boys'' identified per album. We find the largest concentration of ``Bad Boys'' in the album \textit{Speak Now}. This album also has the lowest mean relationship score, as shown in Figure~\ref{fig:relationshipalbum}.}
    \label{fig:badbyalbum}
\end{figure}

We also examined the distribution of ``Bad Boys'' by album, which is shown in Figure~\ref{fig:badbyalbum}. The largest concentration of ``Bad Boys'' is found in \textit{Speak Now}, which is also the album with the lowest relationship score (see Figure~\ref{fig:relationshipalbum}). This correlation may suggest that relationships with ``Bad Boys'' are more likely to end in catastrophe, as a more negative relationship score indicates a more serious/heartbreaking separation between the previous couple after the relationship.

\begin{figure}
    \centering
    \includegraphics[width=\linewidth]{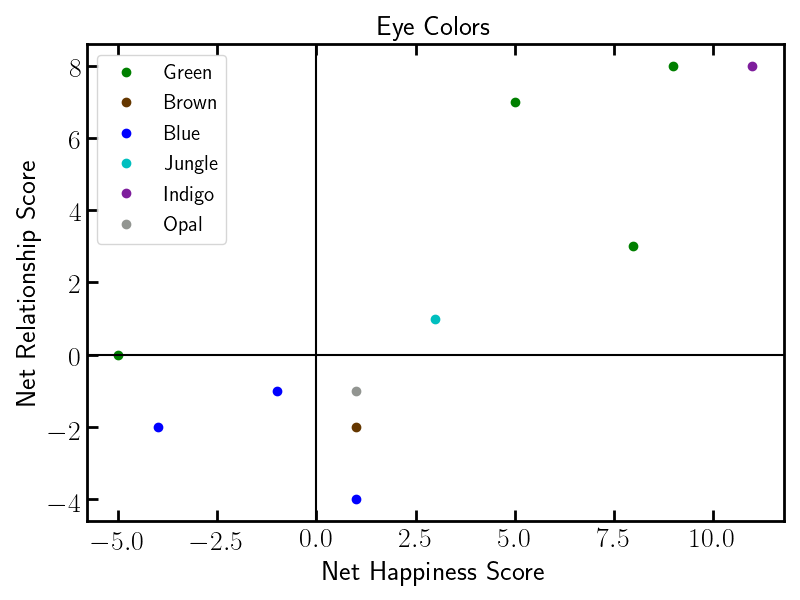}
    \caption{Happiness and relationship scores for songs which mention a specific eye color.}
    \label{fig:eyescatter}
\end{figure}

In addition to searching Swift's songs for occurrences of ``Bad Boys'', we also recorded instances in which the MIQ's eye color was mentioned. Figure~\ref{fig:eyescatter} shows the happiness and relationship scores of the 11 songs in which an eye color was mentioned. Several common eye colors (blue, green, and brown) are occasionally mentioned, along with a few more unique designations. These include the songs \textit{ivy} (``Your opal eyes are all I wish to see''), \textit{I Think He Knows} (``Lyrical smile, indigo eyes''), and \textit{Stay Beautiful} (``Cory's eyes are like a jungle''). We acknowledge that these may simply be poetic ways to refer to more common eye colors - for example, indigo may be a dark blue, while ``jungle-colored'' could refer to some shade of green or hazel. However, to avoid diluting our data with uncertain designations, we consider only the eye colors specifically stated in each song.

\begin{figure}
    \centering
    \includegraphics[width=\linewidth]{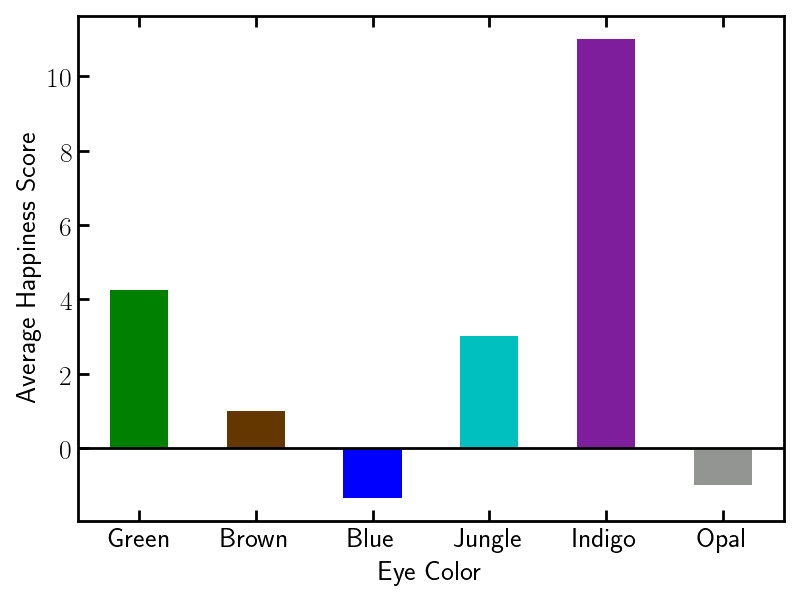}
    \caption{Mean happiness scores by eye color for all songs in which the eye color of the MIQ was mentioned. We find the highest mean score for indigo eyes, while blue eyes show the lowest mean score.}
    \label{fig:eyehappy}
\end{figure}

\begin{figure}
    \centering
    \includegraphics[width=\linewidth]{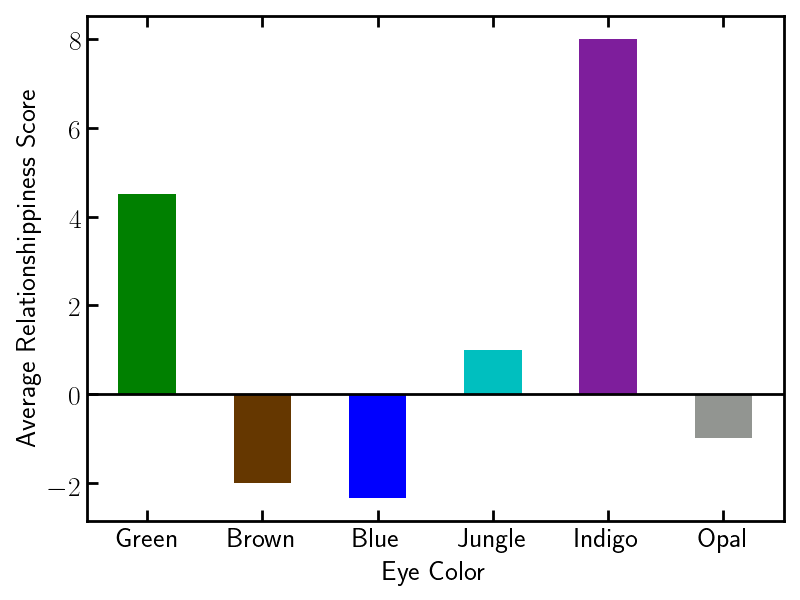}
    \caption{Mean relationship scores by eye color for all songs in which the eye color of the MIQ was mentioned. As in Figure~\ref{fig:eyehappy}, we find the highest mean score for indigo eyes and the lowest mean score for blue eyes.}
    \label{fig:eyerelationship}
\end{figure}

Figures~\ref{fig:eyehappy} and \ref{fig:eyerelationship} show the mean happiness and relationship scores, respectively, for each eye color. We find the lowest mean happiness and relationship scores to occur in songs with blue-eyed MIQs. However, the highest scores by far are found in the single instance of an indigo-eyed MIQ. While we repeat here that these conclusions are based on very small subsets of the data and are thus not statistically significant, our research suggests that indigo- and green-eyed partners may be best for long-term relationships and happiness, while blue eyes can be dreamy but are more likely to lead you astray.

\section{Discussion and Future Work}
\label{sec:conclude}

In this work we presented an analysis of Taylor Swift's full repertoire with the goal of quantifying the representations of relationships and resulting emotions in her songs. Our primary finding is a clear correlation between stronger relationships and more positive emotions. We additionally examine trends in emotional state over time, as well as among subsets of potential partners sorted by perceived reputation and eye color.

As a byproduct of this research, we present the \texttt{taylorswift} code. This code, which is available on GitHub\footnote{https://github.com/meganmansfield/taylorswift}, takes user input on their current relationship status and emotional state and provides suggestions of suitable Taylor Swift songs to match their mood. We anticipate this code being of broad use to the community.

We motivate this work and emphasize that this is one example of a case study. We plan to use an analogous analysis on several different artists, to bring a more comprehensive view of disentangling persona that are presented in the media  with the underlying emotional state of the artists and composers. While this is a useful case study, an extended analysis including the works of several popular artists could provide meaningful and insightful constraints on the differences in perceived persona and emotional psychology. Of particular interest would be to compare the conceived and underlying psychology across genres. 

\section{Acknowledgements}
The authors would like to first acknowledge the great Taylor Swift herself for providing such a rich repertoire of music for analysis. They would also like to thank Cory, Drew, Johnny, Abigail, Stephen, Romeo and Juliet, John, Bobby, Sean, Betty, Inez, James, Rebekah, Bill, Dorothea, Este, Marjorie, and all the other unnamed people who served as inspiration for Swift's songs. M.M. would like to thank the residents of Next House 4E for the late-night discussions that produced the ideas developed in this paper. D.S. is grateful to Gregory Laughlin for useful conversations and suggestions which greatly added to the scientific content of this manuscript. 

\begin{table*}
\centering
\caption{Feelings About Self: one of four happiness criteria}
\label{tab:selffeel}
\begin{tabular}{>{\centering \arraybackslash}p{0.9 cm} >{\centering \arraybackslash}p{6 cm} >{\centering \arraybackslash}p{10 cm}}
\hline \hline
\textbf{Score} & \textbf{Description} & \textbf{Example} \\
\hline
-3 & Feels fully responsible for problems & ``Stupid girl / I should have known / I should have known'' (\textit{White Horse}) \\
-2 & Feels partial responsibility for problems & ``No one teaches you what to do / When a good man hurts you / And you know you hurt him too'' (\textit{happiness}) \\
-1 & Hints at self-deprecation & ``Maybe we got lost in translation, maybe I asked for too much'' (\textit{All Too Well}) \\
0 & No feelings mentioned/ambiguous feelings & \textit{Blank Space} \\
1 & Overall positive with serious insecurities & ``You have pointed out my flaws again / As if I don't already see them / I walk with my head down / Trying to block you out 'cause I'll never impress you'' (\textit{Mean}) \\
2 & Overall positive with some reservations & ``20 questions, we tell the truth / You've been stressed out lately, yeah, me too'' (\textit{It's Nice To Have A Friend}) \\
3 & Secure and trusting in life circumstances & ``And we see you over there on the internet / Comparing all the girls who are killing it / But we figured you out / We all know now / We all got crowns'' (\textit{You Need To Calm Down})\\
\hline \hline
\end{tabular}
\end{table*}

\begin{table*}
\centering
\caption{Glass Half Full: one of four happiness criteria}
\label{tab:glassfull}
\begin{tabular}{>{\centering \arraybackslash}p{0.9 cm} >{\centering \arraybackslash}p{6 cm} >{\centering \arraybackslash}p{10 cm}}
\hline \hline
\textbf{Score} & \textbf{Description} & \textbf{Example} \\
\hline
-3 & All imagery is depressing & ``And then the cold came, the dark days when fear crept into my mind'' (\textit{Back To December}) \\
-2 & Nearly all depressing imagery & ``How you laugh when you lie / You said the gun was mine / Isn't cool, no I don't like you / But I got smarter, I got harder'' (\textit{Look What You Made Me Do}) \\
-1 & Majority depressing imagery & ``Stealing hearts and running off and never saying sorry / But if I'm a thief then / He can join the heist'' (\textit{...Ready For It?}) \\
0 & Equal amounts of happy and sad imagery & ``Rain came pouring down when I was drowning / That's when I could finally breathe'' (\textit{Clean}) \\
1 & Majority positive imagery & ``We're happy, free, confused, and lonely in the best way / It's miserable and magical'' (\textit{22}) \\
2 & Nearly all positive imagery & ``This love is difficult, but it's real / Don't be afraid, we'll make it out of this mess / It's a love story, baby just say yes'' (\textit{Love Story}) \\
3 & All imagery is positive & ``And all I feel in my stomach is butterflies / The beautiful kind, making up for lost time'' (\textit{Everything Has Changed}) \\
\hline \hline
\end{tabular}
\end{table*}

\begin{table*}
\centering
\caption{Stage of Grief (or Lack Thereof): one of four happiness criteria. Note that both Anger and Depression received a score of -3.}
\label{tab:stagegrief}
\begin{tabular}{>{\centering \arraybackslash}p{0.9 cm} >{\centering \arraybackslash}p{6 cm} >{\centering \arraybackslash}p{10 cm}}
\hline \hline
\textbf{Score} & \textbf{Description} & \textbf{Example} \\
\hline
\multirow{2}{0.9 cm}{\centering -3} & Anger & ``Soon she's gonna find / Stealing other people's toys / On the playground won't make you many friends'' (\textit{Better Than Revenge}) \\
& Depression & ``He's the reason for the teardrops on my guitar'' (\textit{Teardrops On My Guitar}) \\
-2 & Bargaining & ``If it's all in my head, tell me now / Tell me I've got it wrong somehow'' (\textit{tolerate it}) \\
-1 & Denial & ``We found wonderland / You and I got lost in it / And we pretended it could last forever'' (\textit{Wonderland}) \\
0 & Acceptance & ``And all the pieces fall / Right into place / ... / So it goes'' (\textit{So It Goes...}) \\
1 & Passively wanting to be happy & ``I wish you would come back / Wish I never hung up the phone like I did'' (\textit{I Wish You Would}) \\
2 & Actively working for her happiness & ``Drop everything now / Meet me in the pouring rain / Kiss me on the sidewalk / Take away the pain'' (\textit{Sparks Fly}) \\
3 & Actively working for her own and others' happiness & ``I'll fight their doubt and give you faith / With this song for you'' (\textit{Ours}) \\
\hline \hline
\end{tabular}
\end{table*}

\begin{table*}
\centering
\caption{A list of all the hyperbolic lines in Taylor Swift's songs. The first two examples (above the horizontal line) are positive hyperbolic lines, and the remaining examples below the line are negative hyperbolic lines.}
\label{tab:hyperbole}
\begin{tabular}{>{\centering \arraybackslash}p{4.5 cm} >{\centering \arraybackslash}p{12 cm}}
\hline \hline
\textbf{Song} & \textbf{Line}\\
\hline
\textit{Starlight} & ``Ooh ooh, he's talking crazy / Ooh, ooh, dancing with me / Ooh, ooh, we could get married / Have ten kids and teach them how to dream'' \\
\textit{Stay Stay Stay} & ``You took the time to memorize me / My fears, my hopes, my dreams, I just like hanging out with you all the time / All those times that you didn't leave / It's been occurring to me I'd like to hang out with you my whole life'' \\
\hline
\textit{Cold As You} & ``You never did give a damn thing honey but I cried, cried for you / And I know you wouldn't have told nobody if I died, died for you'' \\
\textit{Picture To Burn} & ``And if you're missing me you better keep it to yourself / 'Cause coming back around here would be bad for your health'' \\
\textit{Mean} & ``All you are is mean / And a liar / And pathetic / And alone in life, and mean / And mean / And mean / And mean.'' \\
\textit{I Knew You Were Trouble} & ``And the saddest fear comes creeping in / That you never loved me, or her, or anyone, or anything'' \\
\textit{Look What You Made Me Do} & ``I'm sorry, the old Taylor can't come to the phone now. Why? Oh, 'cause she's dead!'' \\
\textit{This Is Why We Can't Have Nice Things} & ``And here's to you 'cause forgiveness is a nice thing to do / I can't even say it with a straight face'' \\
\textit{no body, no crime} & ``Good thing his mistress took out a big life insurance policy / ... / I wasn't letting up until the day he died'' \\
\hline \hline
\end{tabular}
\end{table*}

\begin{table*}[]
\centering
\caption{Seriousness of Topics Discussed: one of four relationship criteria}
\label{tab:seriousness}
\begin{tabular}{>{\centering \arraybackslash}p{0.9 cm} >{\centering \arraybackslash}p{6 cm} >{\centering \arraybackslash}p{10 cm}}
\hline \hline
\textbf{Score} & \textbf{Description} & \textbf{Example} \\
\hline
-3 & Cataclysmic past offenses & ``It rains when you're here and it rains when you're gone / 'Cause I was there when you said forever and always / You didn't mean it baby'' (\textit{Forever \& Always}) \\
-2 & Some past hurt feelings & ``X marks the spot where we fell apart / He poisoned the well, I was lying to myself'' (\textit{Getaway Car})\\
-1 & Unspecified relationship endings & ``And what once was ours is no one's now'' (\textit{Death By A Thousand Cuts}) \\
0 & Not discussed/Pining & ``Please don't be in love with someone else / Please don't have somebody waiting on you'' (\textit{Enchanted}) \\
1 & Puppy love/One night stand & ``With all those nights we're spending / Up on the roof with a school girl crush'' (\textit{King Of My Heart}) \\
2 & Some real world things to discuss & ``But you start to talk about the movies that your family watches / Every single Christmas and I want to talk about that'' (\textit{Begin Again}) \\
3 & Discussion of marriage/equally serious topics & ``And you know that I'd swing with you for the fences / Sit with you in the trenches / Give you my wild, give you a child'' (\textit{peace}) \\
\hline \hline
\end{tabular}
\end{table*}

\begin{table*}
\centering
\caption{Future Prospects: one of four relationship criteria}
\label{tab:future}
\begin{tabular}{>{\centering \arraybackslash}p{0.9 cm} >{\centering \arraybackslash}p{6 cm} >{\centering \arraybackslash}p{10 cm}}
\hline \hline
\textbf{Score} & \textbf{Description} & \textbf{Example} \\
\hline
-3 & Permanent end to communication & ``This time, I'm telling you / We are never, ever, ever getting back together'' (\textit{We Are Never Getting Back Together}) \\
-2 & Significant decrease in contact & ``Say you'll see me again / Even if it's just pretend'' (\textit{Wildest Dreams}) \\
-1 & Possible decrease in contact & ``Guess I'll just stumble on home to my cats / Alone, unless you want to come along'' (\textit{Gorgeous}) \\
0 & No discussion of future/Ambiguous & ``This love is alive back from the dead / These hands had to let it go free but / This love came back to me'' (\textit{This Love}) \\
1 & Casual or potential future plans & ``Hey Stephen, I could give you fifty reasons / Why I should be the one you choose'' (\textit{Hey Stephen}) \\
2 & Some set future plans & ``Morning, his place / Burnt toast, Sunday / You keep his shirt / He keeps his word'' (\textit{You Are In Love}) \\
3 & Marriage/Bound for life & ``I'll be eighty-seven, you'll be eighty-nine / I'll still look at you like the stars that shine'' (\textit{Mary's Song}) \\
\hline \hline
\end{tabular}
\end{table*}

\begin{table*}
\centering
\caption{Declaration of Love from MIQ: one of four relationship criteria. Note that the quote from \textit{ME!} is sung by Brendon Urie and so represents the first-person perspective of the MIQ.}
\label{tab:malefeel}
\begin{tabular}{>{\centering \arraybackslash}p{0.9 cm} >{\centering \arraybackslash}p{6 cm} >{\centering \arraybackslash}p{10 cm}}
\hline \hline
\textbf{Score} & \textbf{Description} & \textbf{Example} \\
\hline 
-3 & He tells all his friends he hates her & ``Did you think I wouldn't hear all the things you said about me?'' (\textit{This Is Why We Can't Have Nice Things}) \\
-2 & He makes a face when her name is mentioned but doesn't publicly hate on her & ``Why'd I have to break what I love so much? / It's on your face, and I'm to blame'' (\textit{Afterglow}) \\
-1 & He doesn't want to date but likes her as a friend & ``And you can't see me wanting you the way you want her / But you are everything to me'' (\textit{Invisible}) \\
0 & No information/Ambiguous & \textit{Dancing With Our Hands Tied} \\
1 & He expressed casual interest in a relationship & ``I say 'I've heard that you've been out and about with some other girl.' / He says 'What you heard is true but I / can't stop thinking about you''' (\textit{Style}) \\
2 & They are dating but not that seriously (she hasn't met his parents) & ``But you pull me in and I'm a little more brave / It's the first kiss'' (\textit{Fearless}) \\
3 & Public declaration of love/commitment & ``I never wanna see you walk away'' (\textit{ME!}) \\
\hline \hline
\end{tabular}
\end{table*}

\begin{table*}
\centering
\caption{Sense of Togetherness: one of four relationship criteria. Note that the ratings between -2 and +2 are based on the lyrics of the full song taken together, so only song titles are given as opposed to specific lyrics.}
\label{tab:together}
\begin{tabular}{>{\centering \arraybackslash}p{0.9 cm} >{\centering \arraybackslash}p{6 cm} >{\centering \arraybackslash}p{10 cm}}
\hline \hline
\textbf{Score} & \textbf{Description} & \textbf{Example} \\
\hline
\multirow{2}{0.9 cm}{\centering -3} & \multirow{2}{6 cm}{\centering Barriers to joint actions} & Distance: ``But you're in London and I break down / 'Cause it's not fair that you're not around'' (\textit{Come Back...Be Here}) \\
 & & Another Woman/Man: ``She said 'James, get in, let's drive' / Those days turned into nights / Slept next to her, but / I dreamt of you'' (\textit{betty}) \\
-2 & No joint actions & \textit{The Man} \\
-1 & More things apart than together & \textit{Holy Ground}, \textit{You Belong With Me} \\
0 & Equal amounts of time together and apart & \textit{willow}, \textit{Treacherous} \\
1 & More things together than apart & \textit{Stay Beautiful}, \textit{New Year's Day} \\
2 & They do everything together & \textit{Paper Rings} \\
3 & No identity as an individual & ``I'm only up when you're not down / Don't wanna fly if you're still on the ground'' (\textit{I'm Only Me When I'm With You}) \\
\hline \hline
\end{tabular}
\end{table*}

\bibliographystyle{aasjournal}

\end{document}